\documentclass[conference]{IEEEtran}

\usepackage{cite}
\usepackage{amsmath,amssymb,amsfonts}
\usepackage{algorithmic}
\usepackage{graphicx}
\usepackage{textcomp}
\usepackage{xcolor}
\usepackage{paralist}
\usepackage{subcaption}
\usepackage{microtype}
\usepackage{hyperref}

\newcommand{\fig}[1]{Fig.~\ref{fig:#1}}

\newcommand{\s}[1]{Section~\ref{sec:#1}}

\newcommand{\fakeparagraph}[1]{\vspace{1mm}\noindent\textbf{#1.}}

\def\BibTeX{{\rm B\kern-.05em{\sc i\kern-.025em b}\kern-.08em
    T\kern-.1667em\lower.7ex\hbox{E}\kern-.125emX}}
\begin{document}

\title{Semi-Automated Design\\of Data-Intensive Architectures}

\author{\IEEEauthorblockN{Arianna Dragoni}
\IEEEauthorblockA{\textit{Politecnico di Milano}\\
Milano, Italy\\
arianna.dragoni@polimi.it}
\and
\IEEEauthorblockN{Alessandro Margara}
\IEEEauthorblockA{\textit{Politecnico di Milano}\\
Milano, Italy\\
alessandro.margara@polimi.it}}

\maketitle

\begin{abstract}
  Today, data guides the decision-making process of most companies.
  Effectively analyzing and manipulating data at scale to extract and exploit
  relevant knowledge is a challenging task, due to data characteristics such
  as its size, the rate at which it changes, and the heterogeneity of formats.
  To address this challenge, software architects resort to build complex
  data-intensive architectures that integrate highly heterogeneous software
  systems, each offering vertically specialized functionalities.
  Designing a suitable architecture for the application at hand is crucial to
  enable high quality of service and efficient exploitation of resources.
  However, the design process entails a series of decisions that demand
  technical expertise and in-depth knowledge of individual systems and their
  synergies.

  To assist software architects in this task, this paper introduces a
  development methodology for data-intensive architectures, which guides
  architects in
  \begin{inparaenum}[(i)]
  \item designing a suitable architecture for their specific application
    scenario, and
  \item selecting an appropriate set of concrete systems to implement the
    application.
  \end{inparaenum} 
  
  To do so, the methodology grounds on
  \begin{inparaenum}[(1)]
  \item a language to precisely define an application scenario in terms of
    characteristics of data and requirements of stakeholders;
  \item an architecture description language for data-intensive architectures;
  \item a classification of systems based on the functionalities they offer
    and their performance trade-offs.
  \end{inparaenum}

  We show that the description languages we adopt can capture the key aspects
  of data-intensive architectures proposed by researchers and practitioners,
  and we validate our methodology by applying it to real-world case studies
  documented in literature.
\end{abstract}

\begin{IEEEkeywords}
software architectures, data-intensive architectures, architecture definition
language, architecture design, design methodology.
\end{IEEEkeywords}

\section{Introduction}
\label{sec:intro}

As data increasingly drive decision-making processes, data management becomes
a key concern for modern organizations, which need to acquire, persist,
analyze, and make data available to internal and external stakeholders based
on their needs.
Over the years, specialized systems have been designed to manage and process
data at scale~\cite{margara:CSur:2023:model}.  Each of them is optimized for
specific tasks but no single system meets all organizational
requirements~\cite{stonebraker:ICDE:05:one_size}.  As a consequence, engineers
resort to build complex \emph{data-intensive architectures} that integrate
multiple such systems to meet the demand of the scenario at hand.

Devising a suitable architecture for a given scenario is challenging, as it
requires careful consideration of numerous factors, including input data
volume, generation frequency, and format, the type of processing to be
performed, the scale of the scenario in terms of number and heterogeneity of
data consumers, just to mention a few.
To mitigate this problem, various reference data-intensive architectures have
been defined in the literature, such as the well-know lambda and kappa
architectures~\cite{davourian:2020:CSur:bd_systems}.  However, engineers still
need to manually select the most suitable architecture for their application
scenario.  Moreover, reference architectures are abstract and high-level, thus
requiring additional work to properly configure them for the requirements of
the company.
In summary, designing a data-intensive architecture for a given application
scenario frequently remains a manual and unstructured process, where the
software architect's decisions often remain implicit and undocumented.

Building on this observation, we use a model-driven engineering
approach~\cite{da2015model} to propose metamodels for describing application
scenarios and architectures.
Furthermore, we propose a methodology to systematize and semi-automate the
definition of data-intensive architectures.
Our methodology works in two steps.
\begin{inparaenum}[(1)]
\item It takes in input the description of the application scenario and
  automatically translates it into a suitable data-intensive architecture.
\item It proposes concrete software systems to implement the architecture and
  provide all the required functionalities.
\end{inparaenum}
By exploiting a rigorous process, the methodology helps engineers in taking
the critical choices behind the definition of data-intensive architectures,
making the motivations behind each choice explicit and well documented, thus
also simplifying future evolution.
To make this possible, we propose formal languages to precisely describe
application scenarios, data-intensive architectures, and distributed data
management and processing systems. 

In summary, our paper brings the following contributions.
\begin{inparaenum}[(1)]
\item It proposes a \emph{scenario description language} that software
  architects can adopt to precisely describe their application scenario,
  including its characteristics and requirements.  Acknowledging the central
  role of data for architectural concerns~\cite{ford:2021:software}, the
  language focuses on data characteristics and data consumers requirements.
\item It introduces an \emph{architecture description language} to describe
  data-intensive architectures.  We show that the language can capture
  reference architectures from the literature and detail the specific way in
  which they are configured to fulfill the scenario requirements.
\item It maps the components of a data-intensive architecture to concrete data
  management and processing systems based on the functionalities and
  guarantees they offer.  This mapping builds on a modeling and classification
  study from the recent literature~\cite{margara:CSur:2023:model}.
\item Based on the above conceptual framework, it introduces a novel
  semi-automated methodology to define a data-intensive architecture starting
  from a given application scenario.
\end{inparaenum}

The paper describes the methodology in details and evaluates its effectiveness
in capturing the key requirements of application scenarios and converting them
to suitable software architectures.

The paper is organized as follows. \s{related} surveys related work.
\s{languages} presents the metamodels we use to define application scenarios and
software architectures. \s{methodology} details our methodology for deriving
data-intensive architectures from application scenarios.  \s{eval} evaluates the
methodology in terms of expressivity, effectiveness, and efficiency.  Finally,
\s{conclusions} concludes the paper indicating future research directions.

\section{Related Work}
\label{sec:related}

Our work crosses the boundaries of various research topics.


At the core of our work lies an Architecture Description Language (ADL)
tailored for data-intensive applications.
ADLs define the components of a software architecture and their
  interrelationships~\cite{medvidovic:TSE:2000:survey_adls,
  malavolta:TSE:2012:survey_adls2}.
Several architectures have been proposed to satisfy the needs of
data-intensive applications, such as the lambda and kappa
architectures~\cite{lin:IntComp:2017:lambda_kappa}.
Few studies have explored ADLs for data-intensive systems, notable examples
being the UML profile in the DICE project~\cite{artac:ICSA:2018:IaC} and the
DAF architecture description
framework~\cite{abughazala:IDSTA:2023:architecture}.
Our ADL builds on a systematic analysis, modeling, and survey of
data-intensive systems~\cite{margara:CSur:2023:model}, encompassing both
system~\cite{kleppmann:2017:designing} and software
engineering~\cite{davourian:2020:CSur:bd_systems} concerns.
The resulting ADL overlaps with the models discussed above, which increases
our confidence on its soundness.
With respect to existing proposals, our work goes beyond the definition of an
ADL and provides a methodology to automatically derive a data-intensive
architecture from an application scenario.


Our Scenario Description Language (SDL) captures the requirements of the
various stakeholders of a data-intensive application.
As acknowledged in recent studies~\cite{davourian:2020:CSur:bd_systems}, the
data-intensive domain lacks established requirement engineering practices.
Yet, the literature in the area clearly points out the central role of data
characteristics to capture the requirements of such
applications~\cite{ford:2021:software}, which are what our language focuses
on.
Related work in the fields target the specification and analysis of
non-functional data requirements.  The interested reader can refer to the
survey in~\cite{davourian:2020:CSur:bd_systems}.
Compared to these proposals, our SDL offers greater granularity by detailing
all processing stages data passes through, moving from sources to consumers.

Our use of data characteristics and quality metrics to guide architectural
decisions draws inspiration from architectural
tactics~\cite{marquez:JSS:2023:architectural_tactics}.
Along this line, future work will explore additional metrics, such as
fault-tolerance and data privacy, currently beyond the scope of our
methodology.

Our definition of scenarios as graphs of transformations draws inspiration
from the dataflow programming model~\cite{akidau:VLDB:2015:dataflow}, widely
adopted in data analytics~\cite{margara:CSur:2023:model}.
Model-driven development approaches for stream processing applications using
dataflow programs have been explored in the
literature~\cite{guerriero:TOSEM:2021:streamgen}.
Our work pursues similar goals, extending them to a broader range of
applications and architectural solutions.


Finally, our work maps architectural components onto concrete systems.  In
doing so, it relies on the vast literature on data-intensive
systems~\cite{kleppmann:2017:designing} and in particular on the model and
survey in~\cite{margara:CSur:2023:model}.
An interesting area of investigation for future work is the automated
generation of configuration and deployment specifications of software systems,
also known as infrastructure-as-code~\cite{artac:ICSA:2018:IaC,
  morris:2020:iac}.
Integrating this with our methodology would streamline the lifecycle of
data-intensive applications, enabling semi-automated deployment from
high-level requirements.

\section{Describing Software Scenarios and Architectures}
\label{sec:languages}

\begin{figure*}[!htb]
  \centerline{\includegraphics[width=.7\textwidth]{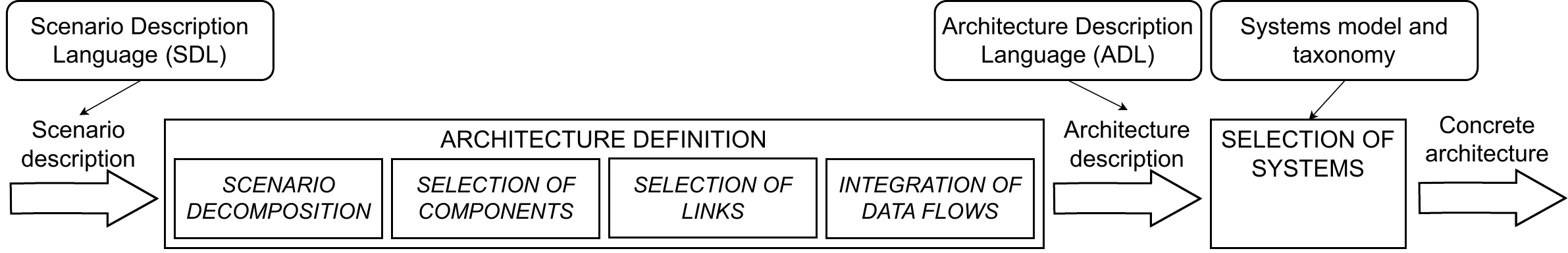}}
  \caption{Overview of the methodology.}
  \label{fig:methodology}
\end{figure*}

This section overviews our methodology (\s{languages:overview}) and presents
the formalism we use to define application scenarios (\s{languages:sdl}),
data-intensive architectures (\s{languages:adl}), and the technologies used to
implement architectural components (\s{languages:systems}).

\subsection{Overview}
\label{sec:languages:overview}

\fig{methodology} provides a visual overview of our proposed methodology.
Engineers define their application scenario by providing a \emph{scenario
  description}.
The methodology works in two steps:
\begin{inparaenum}[(1)]
\item the \emph{architecture definition} step produces an \emph{architecture
  description} consisting of abstract components that communicate with each
  other to realize the overall behavior of the application;
\item the \emph{selection of systems} step proposes data management and
  processing systems to implement the components of the architecture, thus
  realizing a \emph{concrete architecture}.
\end{inparaenum}

We make this possible by introducing formal languages for scenario and
architecture descriptions.
The \emph{scenario description language} (SDL -- \s{languages:sdl}) defines
the main functional and non-functional requirements of a scenario.
The \emph{architecture description language} (ADL -- \s{languages:adl})
specifies abstract components as core building blocks for data-intensive
architectures.
We use an existing \emph{systems model and
  taxonomy}~\cite{margara:CSur:2023:model} to recommend systems for
  implementing each abstract component.

\begin{figure*}[!htb]
  \centerline{\includegraphics[width=.7\textwidth]{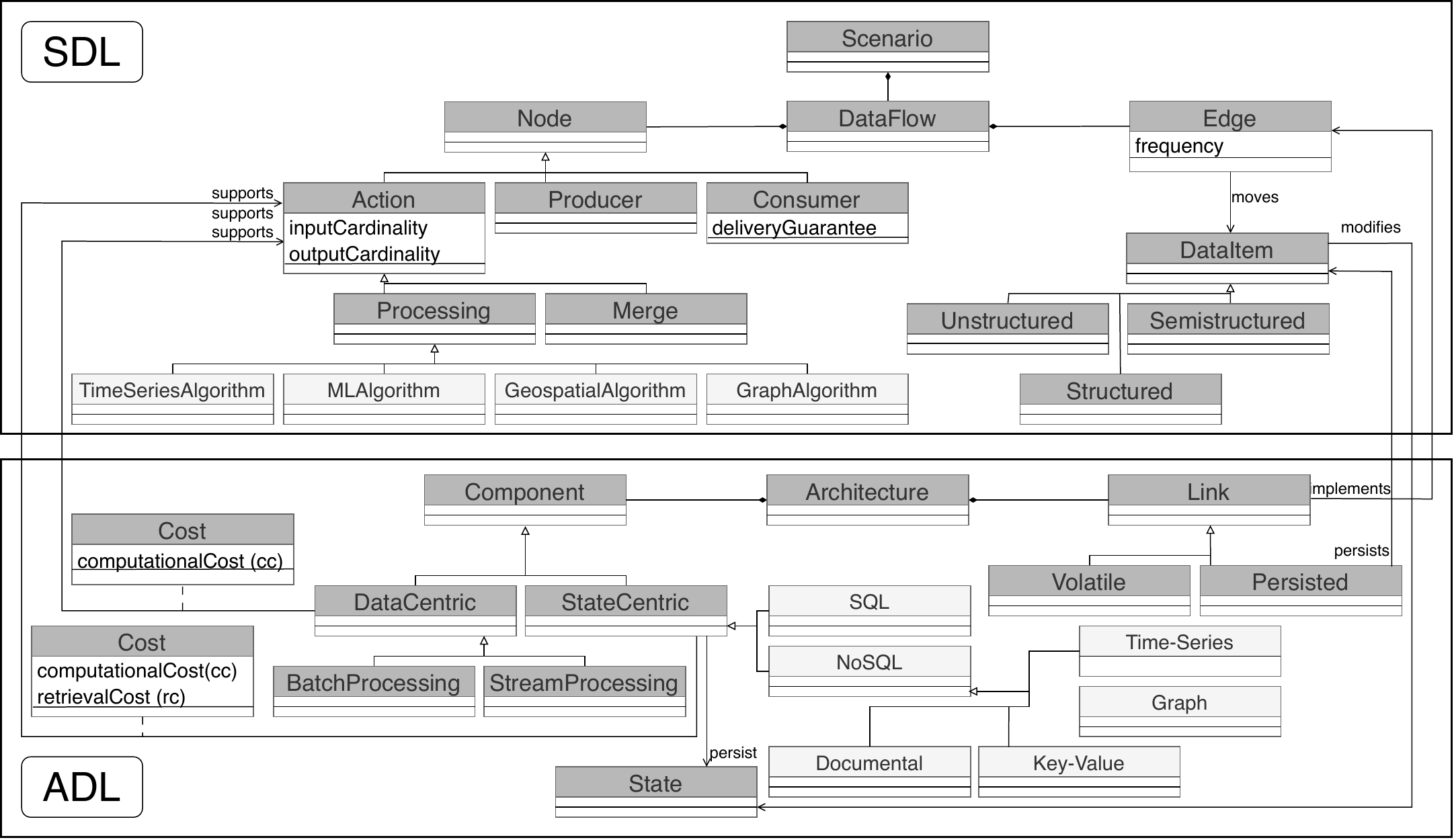}}
  \caption{UML class diagram for the Scenario Description Language (SDL) and
    the Architecture Description Language (ADL).}
  \label{fig:uml}
\end{figure*}

This section presents the SDL and ADL in detail, with their concepts and
relations illustrated in \fig{uml} as a UML diagram.
\s{methodology} details the logic behind the two methodology steps:
architecture definition and selection of systems.

\subsection{Scenario Description Language (SDL)}
\label{sec:languages:sdl}

The SDL (upper part of \fig{uml}) formalizes application scenarios as stages
data transformation that address user needs.

Scenarios are modeled as directed graphs, where \emph{nodes} abstract
operations on data (generation, manipulation, consumption) and \emph{edges}
denote data flows.
We denote the source nodes as \emph{producers} that generate input data for
the application, sink nodes as data \emph{consumers}, and intermediate nodes
as \emph{actions} that transform input data into output data.

Data consists of immutable elements called \emph{data items} that are
transferred from node to node across edges.
Each edge $e$ has an associated \emph{data type} that indicates the common
structure (if any) of all data items traversing $e$.
Data types are categorized as \emph{structured}, \emph{unstructured}, or
\emph{semistructured}.
A structured data type imposes a fixed schema to all data items, meaning that
all data items consist of the same list of typed attributes.  For instance, in
an environmental monitoring application, temperature data items may all be
characterized by a location (of type string), a timestamp (of type long), and
a value (of type float).
Conversely, an unstructured data type does not impose any constraint, as in
the case of free text documents.
Semistructured data types sit in between: they define a structure but they
allow some degree of flexibility in the number of attributes.  For instance
XML and JSON objects may include variable-length lists or maps.
Each edge also has a \emph{frequency} attribute, indicating how often
downstream nodes request data from that edge.

Each producer node in a scenario represents a set of real-world entities that
produce data with similar characteristics, for instance a set of similar
sensors in an environmental monitoring scenario.
Similarly, each consumer node represents a set of users that are interested in
the same results -- that is, they require the same actions to be applied on
input data -- and have the same requirements in terms of frequency of requests
and guarantees on data.
The frequency attribute of a consumer's incoming edge models its request
frequency.
Furthermore, each consumer has an associated \emph{delivery guarantee}
property, which indicates whether the consumer tolerates loss of items (items
are delivered with \emph{at-most once} guarantees), duplicates (\emph{at-least
  once} guarantees), or it requires all results to be delivered as if all
input items were processed once and only once (\emph{exactly once}
guarantees).
For instance, a consumer in an environmental monitoring scenario may be
interested in weekly average of temperatures (result of actions), updated
every second (frequency of requests), and may tolerate loss of data (at-most
once delivery guarantee).

A scenario may involve different consumers, indicating different requirements
for data consumption.  In these cases, we denote as \emph{data flow} for a
consumer $c$ the sub-graph that includes all and only the nodes that are
directly or indirectly connected to $c$ and the edges between these nodes.
Intuitively, a data flow shows the input data and actions needed to meet a
specific customer's requirements.

Actions have associated \emph{input cardinality} and \emph{output cardinality}
attributes, which indicate the amount of data considered and generated by the
action at each evaluation.  These attributes serve as proxies for action
complexity, as further detailed when discussing the computational costs in the
ADL -- cfr \s{languages:adl}.
We consider two types of actions.
\begin{inparaenum}[(1)]
\item \emph{Processing} actions have a single incoming edge.
\item \emph{Merge} actions, by contrast, integrate data coming from multiple
  incoming edges.
\end{inparaenum}
Both processing and merge actions may have one or more outgoing edges.
Processing actions can be further specialized based on the type of data
transformations they define. To ensure adaptability across contexts, we avoid
prescribing a fixed catalog of transformations. Depending on the specific
application domain, transformations may represent time series analysis,
machine learning algorithms, graph algorithms, or other. \fig{uml}
illustrates examples, without aiming for exhaustiveness.

\subsection{Architecture Description Language (ADL)}
\label{sec:languages:adl}

The ADL (lower part of \fig{uml}) formalizes data-intensive architectures as
graphs of software \emph{components} that communicate by exchanging data
items through \emph{links}.
Components model classes of systems with similar characteristics and links
represent technologies to transfer data across systems.

The ADL captures
\begin{inparaenum}[(1)]
\item the capabilities that each software component exposes, and
\item the way in which software components receive and deliver data items.
\end{inparaenum}
Methodologically, we developed the ADL iteratively, starting from both reference
architectures~\cite{davourian:2020:CSur:bd_systems} and real-world use
cases~\cite{real-cases} documented in the literature, and refining it
until we could capture all the key features we extrapolated from
the literature.
We evaluate the modeling capabilities of the language in \s{eval}.
 
Each component \emph{supports} a set of actions it can implement.
For example, a component that supports a time series algorithm is capable of
handling input data in the form of time series and to perform the specified
algorithm.
Similarly, each link \emph{implements} an edge, meaning it implements a way to
transfer data across the actions required in a scenario.
As discussed in \s{methodology}, the first step of our methodology selects
suitable components and links to implement the actions and edges defined in a
scenario.

Links may either persist or not persist data elements traversing them, so we
provide two specialization: \emph{persistent} and \emph{volatile} links.
For example, links that model TCP connections are volatile (they deliver the
data from sender to receiver, but do not store it for future retrieval),
whereas links that model persistent queuing systems such as Apache Kafka or
distributed filesystems such as HDFS enable receiving components to retrieve
data more than once.
The ADL is open to future extensions and refinements through specializations
of these two classes.

Data items traversing links represent domain information, for instance new
purchases performed by the customers of an e-commerce application.
Components transform input data into output data according to the action they
implement.  For instance, a recommender system transforms input data
(customers preferences) into output data (personalized recommendations).
We classify components into two types:
\begin{inparaenum}[(i)]
\item \emph{Data-centric} components apply data transformations either
  continuously, as new data items are received, or upon request, and
  deliver their results through outgoing links.
  \emph{Stream processing} components continuously transform data, as in the
    case of online anomaly detection systems that signal anomaly alerts by
    searching for suspicious patterns in the recent history of received items.
  \emph{Batch processing} components compute data on demand or periodically,
  as in the case of systems that re-train a machine learning model daily,
  using the entire data available when they perform their computation.
\item \emph{State-centric} components store an internal \emph{state}
  representing their view of the application domain.  Upon receiving input
  data items, they update their internal state and make it accessible to
  downstream components through queries.  For instance, a data store may store
  the last 5 purchases of each user and make them accessible upon request.
\end{inparaenum}

Both data-centric and state-centric components perform actions. However,
data-centric components actions produce output data directly, while
state-centric components actions update the component's internal state.
Accordingly, we model the \emph{cost} of performing an action $a$ on a
component $c$ differently for the two cases (cfr \emph{Cost} association
classes in \fig{uml}): data-centric incur a \emph{computational cost} every time
they execute the action, while state-centric component pay an \emph{update
  cost} for updating their state at each execution of an action and a
\emph{retrieval cost} to retrieve information stored in their state and
propagate it to downstream components.
These costs are modeled as functions of the input and output data sizes, as
specified by the action's input and output cardinality attributes.
Currently, architects are expected to define cost functions for the considered
components and actions.
In the future, the adoption of this methodology may lead to a shared library
of cost functions for common components.

\subsection{Systems model and taxonomy}
\label{sec:languages:systems}

Given the evolving landscape of data-intensive systems, we avoid fixed
component categorizations.
In the ADL in \fig{uml}, we follow the established classification of
data-centric components into batch and stream processing systems: this is the
level of granularity at which the current implementation of our methodology
works.
We further show examples of classes of systems that reflect the current state
of the field according to recent work~\cite{margara:CSur:2023:model} (light
gray classes in \fig{uml} -- e.g., time-series or graph processing systems),
and we foresee future evolution and specializations through extensions.
Our methodology is agnostic of the specific characteristics of each class of
systems, and only requires knowledge about the actions that each class (i.e.,
component) supports and an estimate of the cost functions associated to
running a given action on a given component.
Being conceived as a decision support tool, precise estimates of costs are not
necessary, but a good indication of the relative differences between
components can provide useful suggestions to software architects.

\section{Methodology}
\label{sec:methodology}

This section describes the two steps of our methodology -- \emph{architecture
  definition} (\s{methodology:scenario-to-architecture}) and \emph{selection
  of systems} (\s{methodology:selecting-systems}) -- in detail (cfr
\fig{methodology}).

\subsection{Architecture definition}
\label{sec:methodology:scenario-to-architecture}

The first step of the methodology translates a scenario description expressed
in SDL into an architecture description expressed in ADL.
It
\begin{inparaenum}[(i)]
\item decomposes a scenario into its constituting data flows;
\item selects suitable components for each action in each data flow;
\item selects suitable links for each edge in each data flow;
\item merges individual data flows into a single architecture.
\end{inparaenum}

\fakeparagraph{Scenario decomposition}
The methodology decomposes the input scenario into individual data flows,
where each data flow includes all and only the actions that are required to
satisfy the requests of a single consumer.
Nodes shared among multiple data flows are repeated within each of the data
flows they appear in.

\fakeparagraph{Selection of components}
Each data flow is converted into an architecture description, where each node
is translated into a component and each edge is translated into a link.

First, the methodology selects the most suitable type of component to
implement each action.
We formulate this choice as an optimization problem that associates a cost to
each selection of components and aims to minimize the overall cost for each
data flow.

Given a data flow $DF$, we denote the set of its internal nodes (excluding
producers and consumers) as $N$ and the sets of its edges as $E$.
Each internal node $n \in N$ defines an action $a_n$ to be performed on its
input data.
Recall that in a data flow a node can have multiple incoming edges (in the
case of a merge node) and multiple outgoing edges.
We denote the incoming frequency $f^{in}_n$ of a node $n \in N$ as the maximum
of the frequencies associated to its incoming edges, and the outgoing
frequency $f^{out}_n$ of $n$ as the maximum of the frequencies associated to
its outgoing edge.
Intuitively, the incoming frequency represents the maximum frequency at which
input data is requested to upstream components, and the outgoing frequency is
the frequency at which new results are needed for (requested by) downstream
component.

For each node $n \in N$, the optimization problem needs to decide the best
component to implement $n$. 
We consider three macro classes of components: state-centric, data-centric
batch processing, data-centric stream processing.
We assume that the cost functions for implementing action $a_n$ using a given
class of components is known and fixed for any concrete system belonging to
that class and supporting action $a_n$.
As discussed earlier, a coarse-grained estimate of the costs for a given class
of systems is sufficient for the purpose of a decision support tool. However,
our conceptual framework is open to extension through the definition of finer
grained classes of components if needed.

We encode the decision with three Boolean variables $x^{sc}_n$, $x^{dc-b}_n$,
$x^{dc-s}_n$ for each node $n \in N$, where $x^{sc}_n = 1$ indicates that node
$n$ is implemented with a state-centric component, $x^{db-b}_n = 1$ indicates
that node $n$ is implemented with a data-centric batch processing component,
and $x^{db-s}_n = 1$ indicates that node $n$ is implemented with a
data-centric stream processing component.  Each node $n \in N$ is implemented
by exactly one component, as modeled by the following constraint.

\begin{equation*}
\displaystyle \forall_{n \in N}\ \ x^{sc}_n\ +\ x^{db-b}_n\ +\ x^{db-s}_n\ =\ 1 
\end{equation*}

We call $c^{sc}_n$ the cost for implementing action $a_n$ of node $n$ using a
state-centric component.  A state-centric component pays a computational cost
$cc$ for each input data item, which represents the cost of computing the
action and updating the state with the result of the computation, and a
retrieval cost $rc$ for each request coming from downstream components.
We assume $cc$ to be a function of the input data to be used at each
computation, as modeled by the input cardinality attribute $ic_{a_n}$, and
$rc$ to be a function of the output data to be delivered downstream for each
request, as modeled by the output cardinality attribute $out_{a_n}$.
Therefore, its cost depends on the incoming and outgoing frequencies of $n$,
and on the input and output cardinality $ic_{a_n}$ and $out_{a_n}$ as
follows.

\begin{equation*}
  c^{sc}_n\ =\ f^{in}_n \cdot cc(ic_{a_n})\ + \ f^{out}_n \cdot rc(out_{a_n})
\end{equation*}

We call $c^{dc-b}_n$ the cost for implementing action $a_n$ of node $n$ using
a data-centric batch-processing component.  This type of components pay a
computational cost $cc$ every time the computation is triggered (based on the
request of downstream components).  We assume the computational cost to be a
function of the input data to be used at each computation, as modeled by the
input cardinality attribute $ic_{a_n}$ for action $a_n$.
Therefore, the cost $c^{dc-b}_n$ depends on the outgoing frequency of $n$ and
the input cardinality of $a_n$, as follows.

\begin{equation*}
  c^{dc-b}_n\ =\ f^{out}_n \cdot cc(ic_{a_n})
\end{equation*}

We call $c^{dc-s}_n$ the cost for implementing action $a_n$ of node $n$ using
a data-centric stream-processing component.  This type of components pay a
computational cost $cc$ every time a new data item enters the node, which
depends on the input cardinality attribute $ic_{a_n}$ for action $a_n$.
Therefore, the cost $c^{dc-s}_n$ depends on the incoming frequency of $n$ and
the input cardinality of $a_n$ as follows.

\begin{equation*}
  c^{dc-s}_n\ =\ f^{in}_n \cdot cc(ic_{a_n})
\end{equation*}

\noindent
If an action is not supported by a class of components, we assume its cost to
be infinite.

\noindent
The objective of the optimization problem is to minimize the overall cost:

\begin{equation*}
  \min \left(
    \sum_{n \in N}\ \ c^{sc}_n \cdot x^{sc}_n\ +\ 
    c^{dc-b}_n \cdot x^{dc-b}_n\ +\ 
    c^{dc-s}_n \cdot x^{dc-s}_n
  \right)
\end{equation*}

\fakeparagraph{Selection of links}
After selecting the components for implementing each action, the methodology
decides the type of links used to implement edges that transfer data between
components.
As for the case of components, we currently consider two macro-classes of
links, namely persistent and volatile links, where persistent links retain
data items, allowing the outgoing component to retrieve them multiple times.

To decide whether a link is persistent or volatile, we consider the
requirements of the component that consumes data from that link.
Let us denote $C$ the consumer of data flow $DF$, and
$P = \{p^1, \dots, p^n\}$ the set of producers of $DF$.
Let us denote $u_l$ and $d_l$ the upstream and downstream components connected
by link $l$.
The selection algorithm works as follows.

\begin{enumerate}
\item If the upstream component is a producer, that is, $u_l \in P$, and $C$
  requires at least or exactly once delivery guarantees, $l$ is a persistent
  link.  Intuitively, this choice ensures that input data is persisted as soon
  as it enters the system and can be replayed in the case of failures to
  satisfy the requirement of the consumer in terms of delivery guarantees.
\item Otherwise, link $l$ is volatile.  Further cases that require persistent
  links will be considered after integrating individual data flows, as
  detailed later.
\end{enumerate}

\fakeparagraph{Integration of data flows}
After determining components and links for each data flow independently, the
methodology combines the data flows to define a single architecture.

Let us denote $DF$ the set of data flows composing a scenario, and $N$ the set
of nodes in that scenario.
Recall that a single node $n \in N$ may be replicated within multiple data
flows, and for each data flow the methodology has associated a component to
each node based on the specific requirements of that data flow.

Given a node $n \in N$ that is part of a data flow $DF_i \in DF$, we denote
$c_n^i$ the component associated to $n$ in $DF_i$.
Let us define $DF^n = \{DF_1^n, \dots DF_m^n\} \subseteq DF$ the set of data
flows that include node $n \in N$, and $C^n$ the components that implement
node $n$ in the data flows in $DF^n$.

The methodology selects the minimum set of components required to implement
node $n$ as follows:

\begin{enumerate}
\item If $C^n$ contains both a data-centric stream processing component and
  data-centric batch processing component, only the stream processing
  component is preserved.  Indeed, both components compute the same results,
  and stream processing provides stricter guarantees in terms of response time
  for downstream components.
\item If $C^n$ contains a state-centric component, the component is preserved.
  Indeed, the component may be used to store the results of processing and
  make it available to downstream components upon request.
\end{enumerate}

Notice that this procedure enables a single node to be implemented through
multiple components: a data-centric one and a state-centric one.  The
semantics of this case is that the data-centric component is used to perform
the computation, and its results are stored in a state-centric component for
later retrieval.

At the end of this process, the methodology has selected a minimum set of
components $C^n$ to implement each node $n \in N$.
At this point, the methodology decides the links for connecting them.
We preserve any link $l$ connecting an upstream component $u_l$ and a
downstream component $d_l$ in at least one data flow.
We select the persistency of a link $l$ as follows.

\begin{enumerate}
\item A link $l$ connecting a producer to an internal component is persistent
  if it is set as persistent in at least one data flow.  Intuitively, the
  requirement for persistency derives from the delivery guarantees set by
  consumers.
\item If the components associated to the downstream node of $l$ -- that is,
  $C^{d_l}$ -- include a batch processing component, and the set of components
  associated to the upstream node -- that is, $C^{u_l}$ -- does not include a
  state-centric component, then $l$ is persistent.  Intuitively, a batch
  processing component needs to retrieve its input data for each computation.
  This can be done by pulling from the upstream component in the data flow, if
  that component is state-centric, or by accumulating data into a persistent
  link.
\end{enumerate}

\subsection{Selection of systems}
\label{sec:methodology:selecting-systems}

Components and links in architecture descriptions represent generalizations of
concrete technologies that are suitable to satisfy the requirements of a given
scenario.

Moving from components to concrete systems involves aspects that go beyond the
technical characteristics of each system, such as monetary costs or previous
knowledge of specific systems.  Both the set of available systems and the
motivations to favor one system over another may change over time.
For these reasons, our methodology does not select \emph{a single} system for
each component in the architecture, but rather lists the systems that
concretely provide the features defined in the component.
This choice enables our methodology to evolve over time and to specialize if
better categorizations of systems become available for specific application
domains.
Currently, we rely on a taxonomy of systems discussed in recent
literature~\cite{margara:CSur:2023:model}.  This taxonomy guides both the
classifications of components and the list of systems that may implement each
component, which guarantees the consistency between architecture definition
and selection of systems.

Specifically, state-centric and data-centric components represent classes of,
respectively, data management and data processing systems that support the
actions defined in the nodes of a scenario. 
In line with the taxonomy in~\cite{margara:CSur:2023:model}, we consider data
management systems as systems that are primarily designed for the persistence of
data. These systems provide functions for inserting, retrieving, updating and
querying data elements, but do not have the functionality for complex data
processing.
They are further cathegorized depending on the structure of the data they
persist, that is, the specialization of the data item that is moved by the edge
implemented by the incoming link of the data-centric component that represents
the system. 
In particular, engineers select a SQL data management system (e.g.,
MySQL\cite{MySQL}) if the data item is structured, whereas
they select a NoSQL data management system (e.g.,
Redis\cite{Redis},
Cassandra\cite{Cassandra},
MongoDB\cite{MongoDB}) if the data item is unstructured or
semistructured. 
Additional subcategories of NoSQL data management systems can be identified by
further specializing the data item.  For example, if the data items
are documents, NoSQL document databases (e.g., MongoDB) may be a suitable choice.
Data processing systems are those that are primarily designed to perform complex
computations with data. Depending on the specialization of the data-centric
component, different categories of data processing systems can be selected. 
Stream processing systems, which are designed to perform computations
immediately when new data enters the system (e.g.,Flink\cite{Flink}), are well
suited to stream processing components.  On the other hand, batch processing
systems, which aggregate data elements before performing computations (e.g.,
Spark\cite{Spark}), are suitable for batch processing components.

Links refer to the technologies and systems used for data transfer, enabling the
connection of components and meeting the requirements of consumers. 
Volatile links do not represent a category of systems, as their role is merely
to transfer data between components. Instead, they serve as communication
channels that allow two systems to exchange data (e.g., TCP connections).
Persistent links refer to queuing systems (e.g., Kafka \cite{Kafka}) or
distributed file systems (e.g., HDFS \cite{HDFS}), which enable the persistence
of data as it moves from one system to another.

\section{Evaluation}
\label{sec:eval}

In this section, we evaluate our methodology in terms of modeling
capabilities, effectiveness, and efficiency.  Specifically, our evaluation
aims to answer the following three research questions.

\begin{enumerate}[(RQ1)]
\item Is the methodology suitable to model real-world scenarios and
  architectures?
\item Does the methodology guide software architects towards appropriate
  architectures for a given scenario?
\item What is the time required to compute an architecture starting from a
  scenario?
\end{enumerate}

To answer the questions above, we organize our evaluation in three parts.

\begin{enumerate}
\item We analyze four reference architectures for data-intensive applications
  from the literature, and we show that our languages are indeed suitable to
  model the requirements that motivate these architectures (for the SDL) and
  their structure in terms of components and links (for the ADL) --
  \s{eval:rq1}.
\item We rely on a concrete use case discussed in the
  literature~\cite{fb2010}, we model its requirements using our SDL, and we
  show that the methodology generates an architecture description that closely
  resembles the one in the original paper -- \s{eval:rq2}.
\item We execute our methodology on synthetic scenarios of increasing
  complexity, and we show that the methodology provides suggestions within
  tens of seconds even for the most demanding scenarios with hundreds of nodes
  -- \s{eval:rq3}.
\end{enumerate}

The code of our methodology, the scenarios used in the evaluation, and the
scripts used to conduct the experiments are available at
\url{https://github.com/deib-polimi/methodology_simulations.git}.

\subsection{Modeling capabilities}
\label{sec:eval:rq1}

\begin{figure*}[ptb]
  \centering
  \begin{subfigure}{0.24\textwidth}
    \centering
    \includegraphics[width=\textwidth]{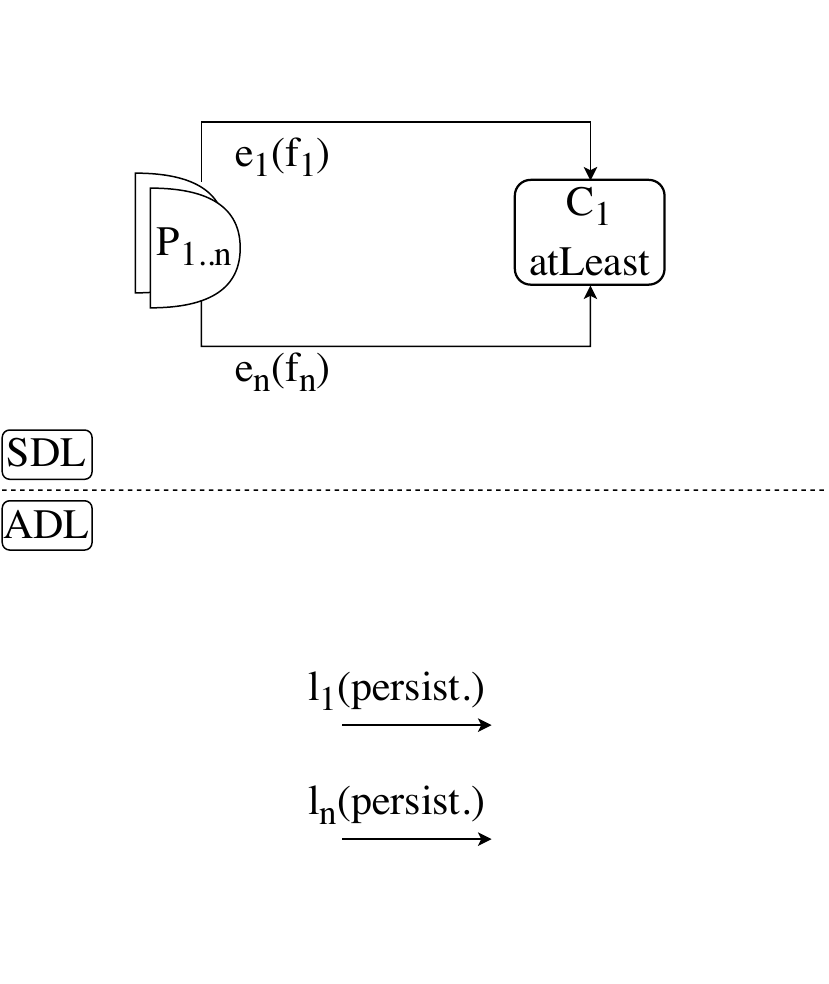}
    \caption{Data Lake}
    \label{fig:data-lake}
  \end{subfigure}
  \hfill
  \begin{subfigure}[b]{0.24\textwidth}
    \centering
    \includegraphics[width=\textwidth]{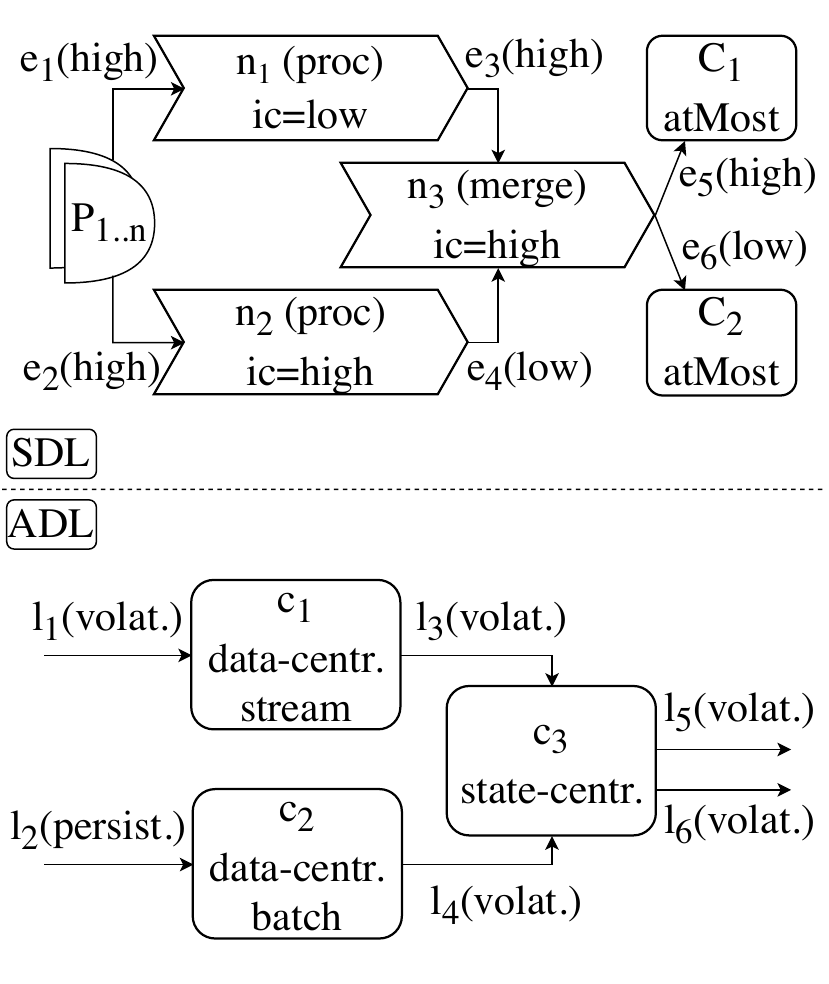}
    \caption{Lambda}
    \label{fig:lambda}
  \end{subfigure}
  \hfill
  \begin{subfigure}[b]{0.24\textwidth}
    \centering
    \includegraphics[width=\textwidth]{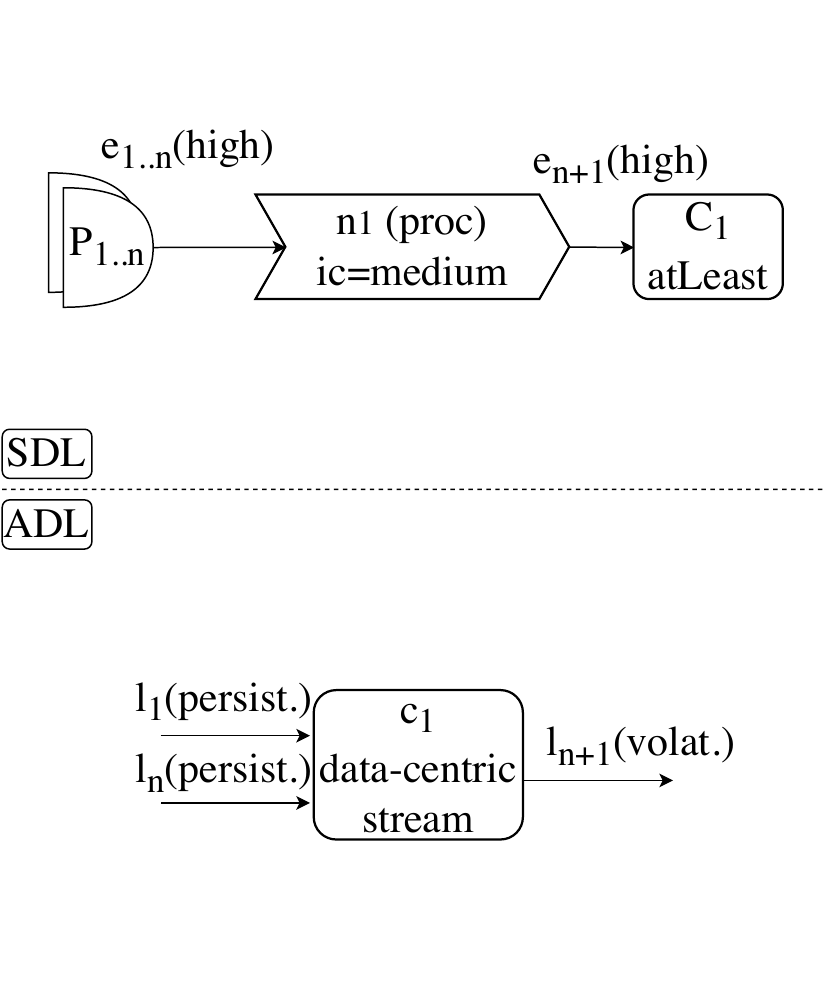}
    \caption{Liquid}
    \label{fig:liquid}
  \end{subfigure}
  \hfill
  \begin{subfigure}[b]{0.24\textwidth}
    \centering
    \includegraphics[width=\textwidth]{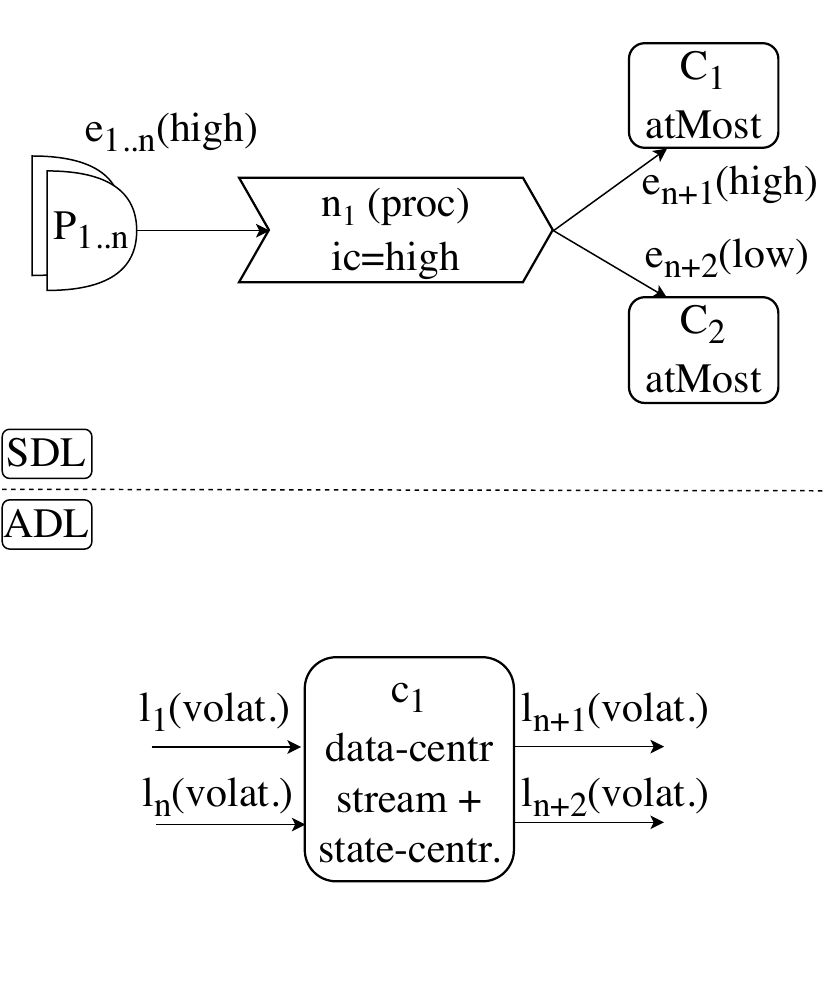}
    \caption{Kappa}
    \label{fig:kappa}
  \end{subfigure}
  \caption{Scenario (top) and architecture (bottom) descriptions for reference
    architectures.}
  \label{fig:eval:reference}
\end{figure*}

To validate the modeling capabilities of our languages, we rely on four
reference architectures that are frequently discussed in the
literature~\cite{davourian:2020:CSur:bd_systems,
  nargesian:VLDB:2019:data-lake}: data lake, liquid, lambda, and kappa.
These are high-level architectural patterns that practitioners have defined to
capture the landscape of data-intensive applications.

We extract from the literature the requirements that guide towards the use of
a given architecture, and we use them to model corresponding scenarios using
our SDL.
We only assign attributes that are relevant for a given architecture, leaving
the others unspecified.
We run our methodology on such scenarios and we verify that we indeed obtain
the reference architectures as presented in the literature.

\fig{eval:reference} illustrates the scenario and architecture descriptions
for the reference architectures.
Next, we report relevant observations that derive from modeling the
architectures within the framework of our methodology.

\fakeparagraph{Data lake} The data lake architecture captures the requirements
of storing data produced at different frequencies and formats from
heterogeneous producers.

We model these requirements using our SDL as shown in the upper part of
\fig{data-lake}.
We represent the producers as $P_1 \dots P_n$, which produce data with
frequencies $f_1 \dots f_n$ over the edges $e_1 \dots e_n$.
The data lake architecture does not prescribe any specific processing to be
performed on such data.  
Accordingly, in our SDL, we model the processing logic by means of a single
consumer ($C_1$) that requires data to be persisted (at least once delivery
guarantee).

Our methodology produces the architecture description shown in the lower part
of \fig{data-lake}.  It correctly identifies the need for storing input data
as it enters the system, and models this need using persistent ingestion links
-- $l_1 \dots l_n$ in \fig{data-lake}.

\fakeparagraph{Lambda} The lambda architecture captures the requirements of
performing different types of computations on input data.  These computations
range from lightweight processing of high-frequency data to heavy
computations on large volumes of data (e.g., historical data) that cannot be
easily performed in an incremental fashion.

We model these requirements in our SDL (upper part of \fig{lambda}) using two
processing nodes $n_1$ and $n_2$, which consume different volumes of data at
each evaluation, as modeled through different input cardinality $ic$ in
\fig{lambda} (indicated as $low$ for $n_1$ and $high$ for $n_2$).
For simplicity, we omit the output cardinality in all the actions in
\fig{eval:reference}, and we always assume it to be low, meaning that the
volume of data is reduced after performing an action.
The results of $n_1$ and $n_2$ are delivered to consumers.  The need for
integrating these results is modeled through a merge node -- $n_3$ in
\fig{lambda}.
Consumers request output data with different requirements in terms of
frequency.  We model them using $C_1$ and $C_2$ in \fig{lambda}, where $C_1$
consumes data with high frequency through edge $e_5$ and $C_2$ consumes data
with low frequency through edge $e_6$.

Our methodology produces the architecture description shown in the lower part
of \fig{lambda}.  It suggests using two separate data-centric components, a
stream processing one ($c_1$ in \fig{lambda}) for lightweight computations and
a batch processing one ($c_2$ in \fig{lambda}) for heavy computations.
Regardless of the delivery guarantees of producers, it always suggests
persisting data before ingesting it into the batch processing component.
Furthermore, the architecture persists the results of heavy computations
within a state-centric merge component, allowing consumers to retrieve them on
demand.

\fakeparagraph{Liquid} The liquid architecture presents two differences with
respect to the lambda architecture in terms of requirements:
\begin{inparaenum}[(i)]
\item Heavy computations can be easily made incremental, thus reducing their
  processing cost upon receiving a new input data item.  We model this
  difference by exploiting a single processing node $n_1$, without
  differentiating the computational costs of different operations -- modeled
  through a medium input cardinality $ic$ -- cfr \fig{liquid}.
\item All consumers request results at high frequency.  We model this
  requirement through a single consumer $C_1$ in \fig{liquid}.
\end{inparaenum}

Our methodology produces the architecture description shown in the lower part
of \fig{liquid}.  It captures the differences with respect to the lambda
architecture and proposes a data-centric stream processing component for all
input data -- \fig{liquid}.

\fakeparagraph{Kappa} The kappa architecture serves consumers with different
needs, which we model in the scenario description in the upper part of
\fig{kappa} using consumers $C_1$ and $C_2$ with different input frequencies
(high and low, respectively).
Computations on input data may become expensive as the input data grows, but
the typical adoption of this architecture reduces the amount of data to be
considered (e.g., by limiting the temporal range under analysis for
computations on historical data).
In this way, the computational cost remains similar across operations.  In our
scenario description, we model this assumption through a single processing
node $n_1$.

Our methodology produces the architecture description shown in the lower part
of \fig{kappa}.  It captures the different requirements of the consumers by
suggesting two different components for node $n_1$ -- a data-centric stream
processing one and a state-centric one.  The state-centric component stores
the results of stream processing computations, making them available on demand
to consumers with low request frequency.

\subsection{Effectiveness}
\label{sec:eval:rq2}

\begin{figure}[pb]
  \centering
  \includegraphics[width=0.7\columnwidth]{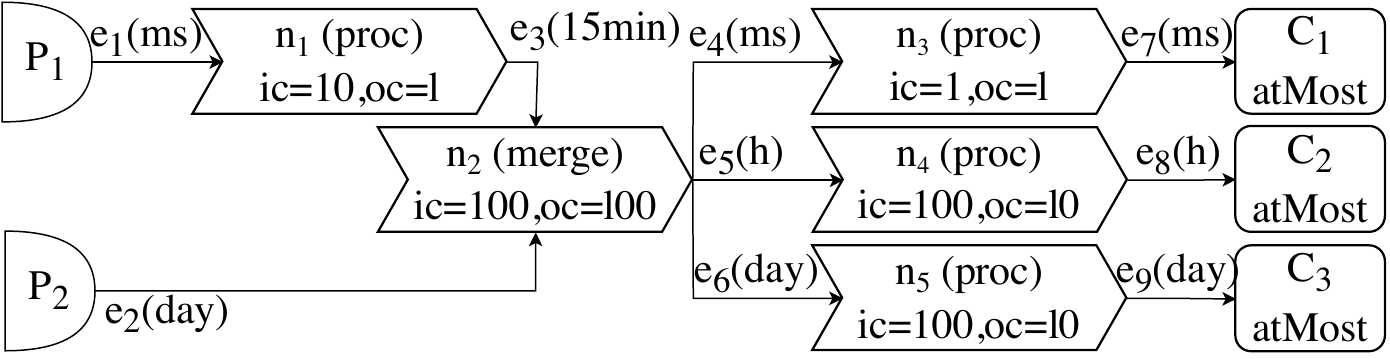}
  \caption{Facebook use case: scenario description.}
  \label{fig:fb-sdl-eval}
\end{figure}

To measure the effectiveness of our methodology, we rely on the concrete use
case discussed in~\cite{fb2010}, which describes the requirements and the
internal architecture of Facebook.

\fig{fb-sdl-eval} illustrates the scenario description we extrapolated based
on the information in~\cite{fb2010}.
We identify two producers:
$P_1$, which continuously delivers event logs and clicks from users, and
$P_2$, which delivers users information daily.

A processing node $n_1$ aggregates logs from $P_1$.  Based on information from
the application domain, we assume that $n_1$ receives data with a rate in the
order of milliseconds (edge $e_1$ in \fig{fb-sdl-eval}), and the aggregation
is performed on groups of tens of input data items that reduced to a single
output data item (that is, the input cardinality $ic$ for node $n_1$ is 10 and
its output cardinality $oc$ is 1).

A merge node $n_2$ processes data coming from $n_1$ and $P_2$: as stated
in~\cite{fb2010}, logs from $n_1$ are evaluated at intervals of 5--15 minutes,
and user data from $P_2$ at daily intervals.
We assume the merge action to consume and output large volumes of data. In the
scenario in \fig{fb-sdl-eval}, we model this assumption by assigning an input
cardinality parameter to $n_2$ that is 10 times larger than the input
cardinality of $n_1$ and an output cardinality parameter that is equal to the
input cardinality.

According to~\cite{fb2010}, there are three classes of users -- that is,
consumers in our SDL, which we model with consumers $C_1$, $C_2$, $C_3$:
$C_1$ demands simple but frequent computations (we assume a rate of requests in the
order of milliseconds),
$C_2$ and $C_3$ demand complex computations with a rate of hours or days,
respectively.
We model the computations required by each of these consumers with three
additional processing nodes:
$n_3$ processes data for consumer $C_1$ -- we model the simplicity of the
processing task by assigning an input and output cardinality of 1;
$n_4$ and $n_5$ process data for consumers $C_2$ and $C_3$, respectively -- we
model the complexity of these tasks by assigning an input cardinality of 100
and an output cardinality of 10.
Although this is not specified in~\cite{fb2010}, we assume consumers to
tolerate loss of data.  As discussed later, this choice does not affect the
results obtained with our methodology.

Since there is no detailed information on the specific actions performed by
nodes, we assumed computational cost functions to be linearly proportional to
the input cardinality and to be identical for all systems.  Alternative
choices may affect the final result.
Finally, from~\cite{fb2010}, we know that data items traversing all edges are
structured, and we omit it for simplicity in \fig{fb-sdl-eval}.

\begin{figure*}[htbp]
  \centering
  \begin{subfigure}[b]{0.3\textwidth}
    \centering
    \includegraphics[width=\textwidth]{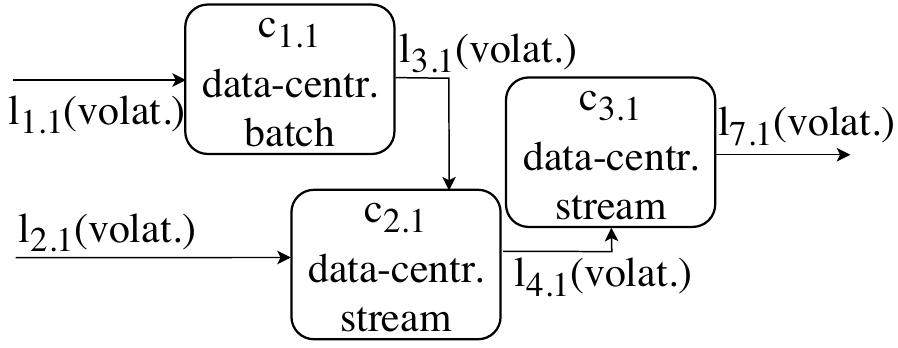}
    \caption{$DF^1$}
    \label{fig:df1-fb}
  \end{subfigure}
  \hfill
  \begin{subfigure}[b]{0.3\textwidth}
    \centering
    \includegraphics[width=\textwidth]{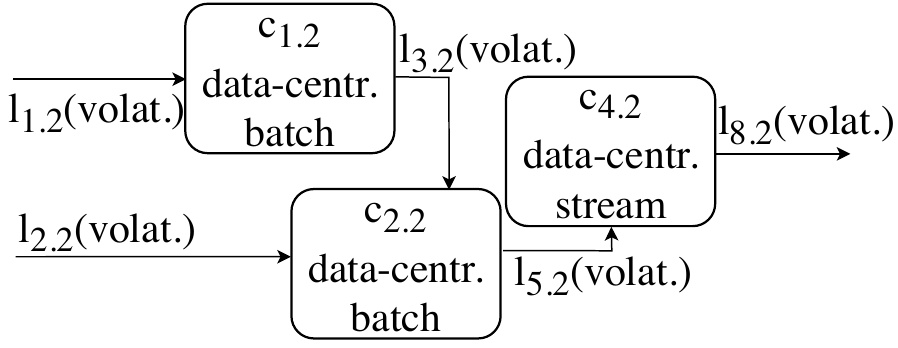}
    \caption{$DF^2$}
    \label{fig:df2-fb}
  \end{subfigure}
  \hfill
  \begin{subfigure}[b]{0.3\textwidth}
    \centering
    \includegraphics[width=\textwidth]{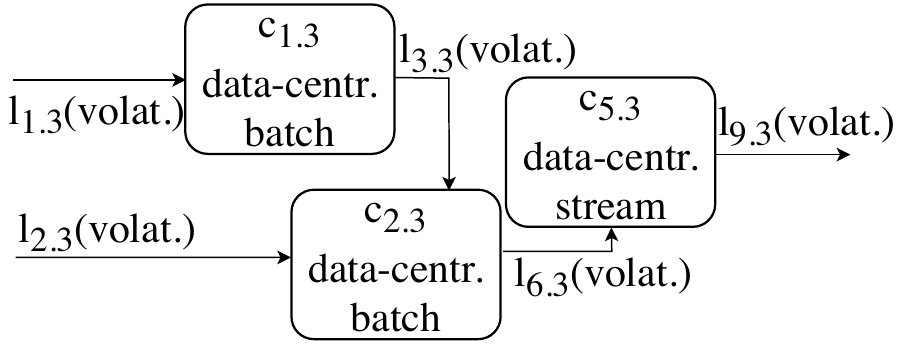}
    \caption{$DF^3$}
    \label{fig:df3-fb}
  \end{subfigure}
  \caption{Facebook use case: components for each data flows.}
  \label{fig:fb-dfs}
\end{figure*}


Starting from the scenario description in \fig{fb-sdl-eval}, our methodology
extracts three data flows $DF^1$, $DF^2$, $DF^3$ for consumers $C_1$, $C_2$,
and $C_3$, respectively, and assigns them the components shown in \fig{fb-dfs}.
Each data flow includes a component for node $n_1$ and one for node $n_2$,
which are required by all three consumers.  Node $n_1$ is associated with a
data-centric batch processing component in all data flows.  Indeed, it
receives data at a high frequency, but it produces output data periodically,
every few minutes.
Node $n_2$ is associated to a data-centric stream processing component in
$DF^1$ and to a data-centric batch processing component in $DF^2$ and $DF^3$,
due to the heterogeneous rates of requests coming from the consumers.
Finally, $n_3$, $n_4$ and $n_5$ are associated to data-centric components.
For each of these components, the rate of output requests is equal to the rate
of input data.  Thus, the choice of a stream or batch processing component
depends on the computational cost function associated to the specific action
for each of the two technologies.  Unfortunately, this information cannot be
easily extrapolated from~\cite{fb2010}, so we assign identical cost functions,
in which case our methodology suggests a stream processing component to
minimize the latency for delivering results.

\begin{figure}[ptb]
  \centering
  \includegraphics[width=0.7\columnwidth]{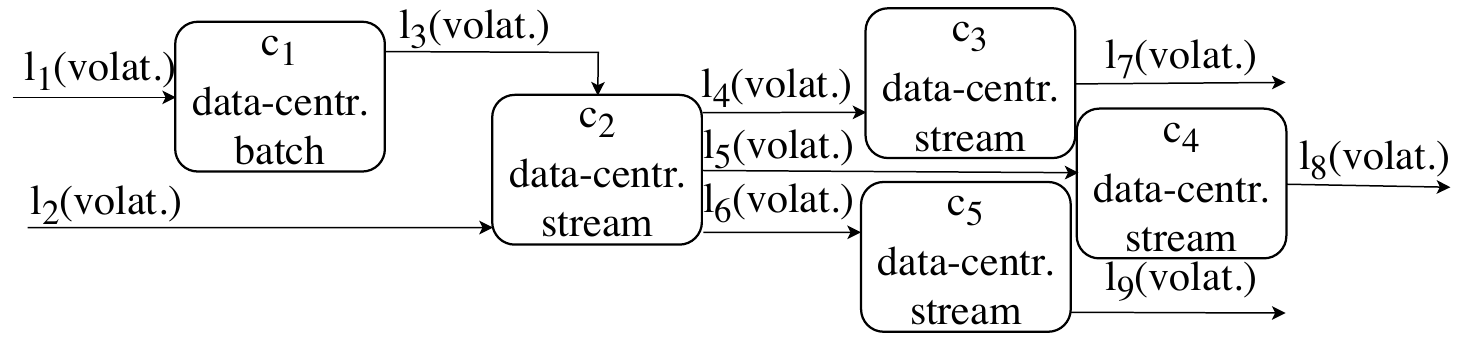}
  \caption{Facebook use case: architecture description.}
  \label{fig:adl-fb}
\end{figure}

The final architecture after merging the components of the three data flows is
shown in \fig{adl-fb}.  Notice in particular that a stream processing
component is selected for node $n_2$, as a result of merging batch and stream
processing components from different data flows.

The architecture obtained by running our methodology is structurally identical
to the one described in the original paper~\cite{fb2010}.
This confirms that our methodology guides towards architectural choices that
are consistent with the ones described for the use case.
In terms of technologies, the paper adopts batch processing technologies for
data-centric components due to the systems available at that time (when stream
processing components were not available).  The actual choice of the classes
of systems to adopt for each component depends on the cost functions the
efficiency of each class in implementing a given action, which varies as new
technologies become available.

\subsection{Efficiency}
\label{sec:eval:rq3}

\begin{figure}[ptb]
  \centering
  \begin{subfigure}{0.25\textwidth}
    \centering
    \includegraphics[width=\textwidth]{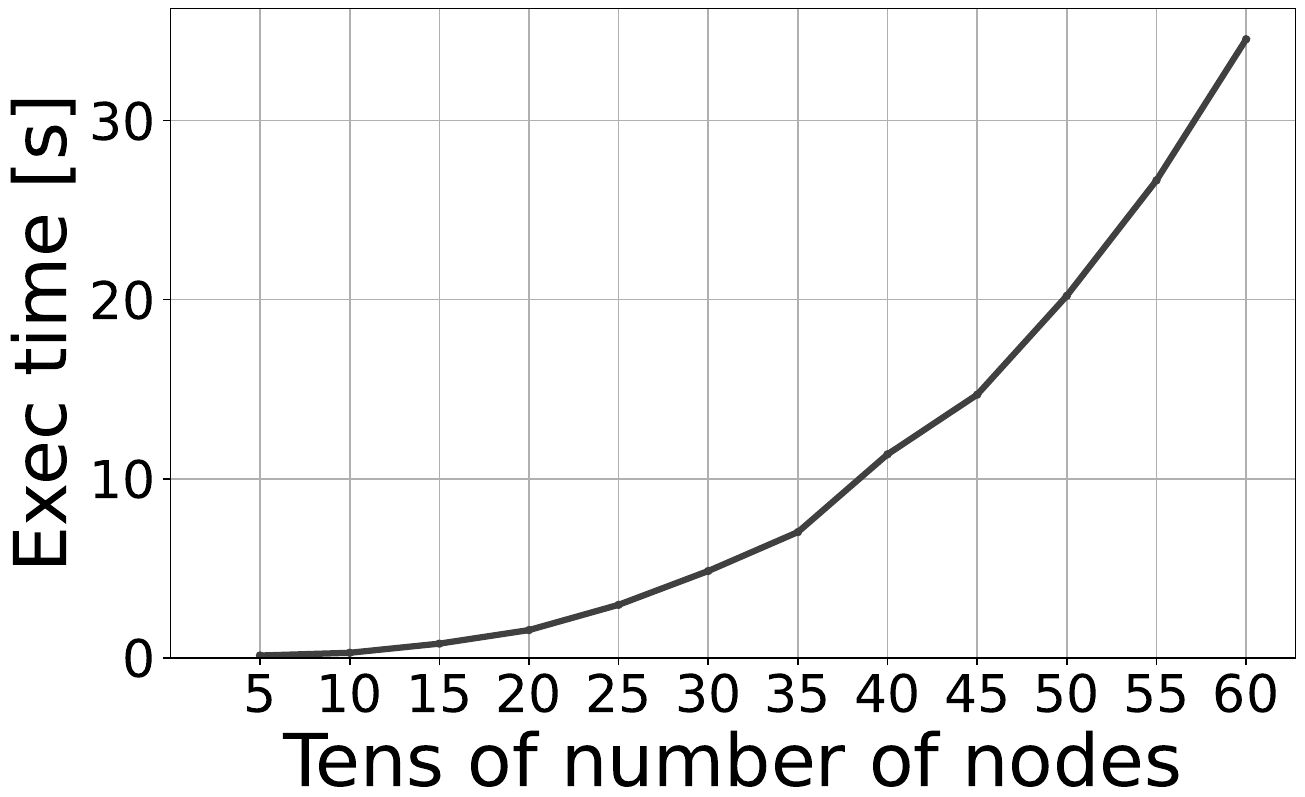}
    \caption{Single data flow}
    \label{fig:single-df}
  \end{subfigure}
  \\
  \begin{subfigure}{0.25\textwidth}
    \centering
    \includegraphics[width=\textwidth]{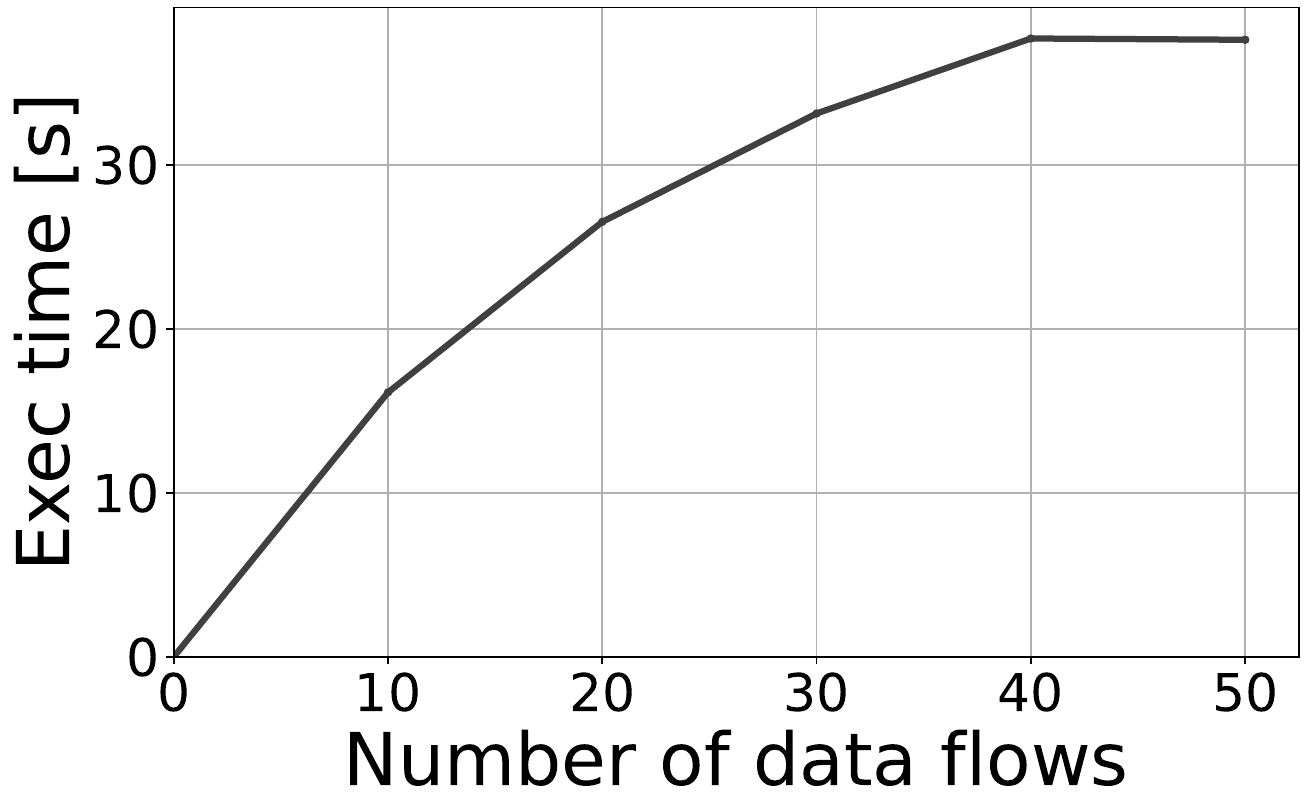}
    \caption{Multiple data flows (300 nodes)}
    \label{fig:multi-df}
  \end{subfigure}
  \caption{Execution time of the methodology.}
\end {figure}

To evaluate the execution time of our methodology, we rely on synthetic
scenarios of increasing complexity, and we run the entire methodology.
Our methodology is written in Python.  To solve the linear optimization
problem, we rely on the PuLP library
version~2.8\footnote{\url{https://coin-or.github.io/pulp/}}.
We execute all the experiments on a M3 MacBook Pro with 24GB of RAM running
MacOS~15.0.1, and using Python~3.11.9.  We run each experiment three times, and
we report the average value.

In generating synthetic scenarios, we consider two cases:
\begin{inparaenum}[(i)]
\item We generate a single data flow and increase the number of nodes in that
  data flow;
\item We consider an increasing number of data flows, with a fixed number of
  nodes.
\end{inparaenum}
Intuitively, the first case evaluates the complexity of the optimization
problem, whereas the second case evaluates the complexity of the data flow
integration problem.
We report the processing times we measured for the two cases in
\fig{single-df} and \fig{multi-df}, respectively.

The results show that our methodology can provide results within seconds of
execution even for (unrealistically) demanding scenarios of hundreds of nodes
or hundreds of consumers.  This demonstrates that the methodology is suitable
as an interactive decision support tool, where software architects may even
step-wise refine their assumptions on the scenario to validate the impact on
the suggested architecture.

\subsection{Discussion and threats to validity}
\label{sec:eval:discussion}

\fakeparagraph{Discussion} The results we measured in the previous sections
lead us to conclude that our methodology is indeed capable of capturing the
key requirements of data-intensive scenarios and the recurring patterns of
data-intensive architectures.  Moreover, it suggests appropriate architectures
for the scenarios at hand.  The execution times we measured testify that the
methodology can be used as an interactive decision support tool even in the
presence of complex scenarios.

\fakeparagraph{Internal threats} In the effectiveness validation, we
extrapolated some parameters of the Facebook architecture, but we did so in a
manner that appears realistic given the studied scenario.  Furthermore, the
difference between the data volumes and rates at play is such that we are
confident the methodology would not vary for minimal parameter changes.  The
selection of actual systems depends on the cost of executing the specific
actions on the available technologies, which is out of the scope of our
evaluation.

\fakeparagraph{External threats} The case studies we selected may not be
representative of the entirety of cases, but we drew them from literature
surveys aiming to be comprehensive.

\fakeparagraph{Construct and conclusion threats} We modeled scenarios based on
our understanding of scenarios and architectures, but the terminology in the
papers used is clear and allows us to be confident in the correct modeling.

\section{Conclusions}
\label{sec:conclusions}

This paper introduced a methodology to support software architects in
designing complex data-intensive architectures.
The methodology starts from a description of the application scenario at hand
in terms of data characteristics and requirements of stakeholders.
It produces an architecture consisting of abstract components, and it further
suggests concrete software systems that may implement each of these
components.

As software systems are increasingly built as compositions of
vertically-specialized services, architectural decisions become even more
important to realize solutions that efficiently fulfill the requirements of
stakeholders.
In this context, our methodology offers a systematic way to document and
evaluate architectural design decisions.
Confident in the potential of our work, we plan to extend it along several
directions. These include broadening the methodology's scope from
data-intensive systems but also operational systems, along with exploring the
interplay between these two domains. Additionally, we aim to develop tools to
analyze the impact of architectural changes during software maintenance.
Lastly, we intend to create a detailed catalog of architectural tactics,
providing more targeted guidance for the selection and configuration of
systems.

\section*{Acknowledgements}

\noindent
We acknowledge financial support from the PNRR MUR project
PE0000021-E63C22002160007-NEST.

\end{document}